%% file: main.tex
\documentclass{acmsmall}

\input{mystyle}

 \newcommand{\mytitle}{Towards a Compiler for Reals}

 \title{\mytitle}
 \author{EVA DARULOVA and VIKTOR KUNCAK\footnote{This work is supported in part by the European Research Council (ERC) project ``Implicit Programming''.}
     \affil{Ecole Polytechnique Federale de Lausanne}}

 \begin{abstract}
 \input{abstract}
 \end{abstract}

\begin{document}


 \maketitle

 \input{introduction}

 \input{overview}

 \input{problem_def}

 \input{ranges}

 \input{straightline}

 \input{loops}

 \input{discontinuities}

 \input{related}

 \input{conclusion}

\bibliographystyle{ACM-Reference-Format-Journals}
\bibliography{main}

 \end{document}

%% file: mystyle.tex
\usepackage{amsmath}
\usepackage{amssymb}
\usepackage{listings}
\usepackage[usenames,dvipsnames,svgnames,table]{xcolor}
\usepackage{hyperref}
\usepackage{booktabs}  

\usepackage[T1]{fontenc}  
\usepackage[scaled=0.80]{beramono}  
\usepackage[letterspace=-60]{microtype}

\lstdefinelanguage{pseudo}{
  alsoletter={@,=,>},
  morekeywords={
  abstract, case, catch, class, def, do, else, extends, false, final, finally,
  for, if, implicit, import, match, new, null, object, override, package,
  private, protected, requires, return, sealed, super, this, throw, trait, try,
  true, type, val, var, while, yield, domain, 
  assert, forAll,  _, return, @generator, ensure, require,
  ensuring,=>, 
  assertBound, jacobian, derivative, Input, Output},
  sensitive=true,
  morecomment=[l]{//},
  morecomment=[s]{/*}{*/},
  commentstyle=\color{gray},
  showstringspaces=false,
  columns=fullflexible,
  mathescape=true,
  numberstyle=\tiny,
  basicstyle=\small\ttfamily, 
  numbers=left,                   
  morestring=[b]"
}

\lstdefinelanguage{scala}{
  alsoletter={@=>},
  morekeywords={
  abstract, case, catch, class, def, do, else, extends, false, final, finally,
  for, if, implicit, import, match, new, null, object, override, package,
  private, protected, requires, return, sealed, super, this, throw, trait, try,
  true, type, val, var, while, with, yield, domain, postcondition,
  precondition,invariant, constraint, assert, forAll,  _, return, @generator,
  ensure, require, ensuring,=>, Real, certainly, possibly, certify, errorBound,
  assertBound
  },
  sensitive=true,
  morecomment=[l]{//},
  morecomment=[s]{/*}{*/},
  commentstyle=\color{gray},
  showstringspaces=false,
  columns=fullflexible,
  mathescape=true,
  numberstyle=\tiny,
  basicstyle=\small\ttfamily,
  numbersep=5pt,
  stepnumber=2,
  numbers=none,                   
  morestring=[b]"
}
\newcommand{\code}[1] {\lstinline!#1!}

\newcommand{\tl}{\tilde}
\newcommand{\tlf}{\tilde{f}}
\newcommand{\tlx}{\tilde{x}}
\newcommand*\norm[1]{ \lvert #1 \rvert }
\newcommand{\R}{\mathbb{R}}
\newcommand*\nNorm[1]{ \left\lvert #1 \right\rvert}
\newcommand*\colvec[3]{  \begin{pmatrix}#1\\#2\\#3\end{pmatrix}}
\newcommand{\fs}{f_*}

\newcommand{\eqn}[1]{
 \begin{align*}
  #1
 \end{align*}}

\newcommand{\eqnnum}[2]{ 
 \begin{align}
 #1
 #2
 \end{align}}

\newcommand{\sparagraph}[1]{\smallskip\noindent \emph{#1}}
\newcommand*\partialDer[2]{ \frac{\partial #1}{\partial #2}}

%% file: abstract.tex
Numerical software, common in scientific computing or embedded systems,
inevitably uses an approximation of the real arithmetic in which most
algorithms are designed. In many domains, roundoff errors are not the only
source of inaccuracy and measurement and other input errors further
increase the uncertainty of the computed results. Adequate tools are needed to
help users select suitable approximations, especially for safety-critical
applications.

We present the source-to-source compiler Rosa which takes as input a
real-valued program with error specifications and synthesizes code over an
appropriate floating-point or fixed-point data type. The main challenge of
such a compiler is a fully automated, sound and yet accurate enough numerical
error estimation. We present a unified technique for floating-point and
fixed-point arithmetic of various precisions which can handle nonlinear arithmetic,
determine closed- form symbolic invariants for unbounded loops and quantify
the effects of discontinuities on numerical errors. We evaluate Rosa on a
number of benchmarks from scientific computing and embedded systems and,
comparing it to state-of-the-art in automated error estimation, show it
presents an interesting trade-off between accuracy and performance.

%% file: introduction.tex
\section{Introduction}

Numerical software, common in scientific computing and embedded systems,
inevitably approximates the real arithmetic in which its algorithms are
typically designed. In addition to roundoff errors from finite-precision
arithmetics such as floating-point or fixed-point, many problem domains come
with additional sources of imprecision, such as measurement and truncation
errors, increasing the uncertainty on the computed results. We need adequate
tools to help developers understand whether the computed values meet the
accuracy requirements and remain meaningful in the presence of the errors.
This is particularly important for safety-critical systems.

Today, however, accuracy in numerical computations is often an afterthought.
We write programs in a user-selected finite-precision data type and then,
perhaps, try to verify that it meets our expectations. This is problematic for
several reasons. Firstly, it introduces a mismatch between the real-valued
algorithms we write on paper and the low-level implementation details of
floating-point or fixed-point arithmetic. Secondly, finite-precision source code
semantics prevents the compiler from applying many optimizations (soundly),
as, for instance, associativity no longer holds. And lastly, numerical errors
remain implicit if they are only the result of some separate analysis tool.

We propose a different strategy. The programmer writes the program in a
real-valued specification language and makes numerical errors explicit in pre- and
postconditions. It is then up to our compiler to determine an appropriate data
type which fulfills the specification but is as efficient as possible and to
generate the corresponding code. This is in particular attractive for fixed-point
arithmetic, often used in embedded systems, where the code generation
can be quite challenging as the programmer has to ensure (implicit) decimal
points are aligned correctly.

Clearly, one of the key challenges of such a compiler is to determine how
close a finite-precision representation is to its ideal implementation in real
numbers. While techniques exist which can handle linear operations
successfully~\cite{Goubault2011,Darulova2011}, precise and sound error
estimation remains difficult in the presence of nonlinear arithmetic. Roundoff
errors and error propagation depend on the ranges of variables in complex and
non-obvious ways; even determining these ranges precisely for nonlinear code
poses a challenge. Furthermore, due to numerical errors, the control flow in
the finite-precision implementation may diverge from the ideal real-valued
one, taking a different branch and producing a result that is far off the
expected one. Quantifying discontinuity errors is hard due to many
correlations and nonlinearity but also due to lack of smoothness or continuity
of the underlying functions that arise in practice \cite{Chaudhuri2011}. In
loops, roundoff errors grow, in general, unboundedly. Even if an iteration
bound is known, loop unrolling approaches scale poorly when applied to
nonlinear code.

We have addressed these challenges and present here our results towards the goal
of a verifying compiler for real arithmetic. In particular, we present
\begin{itemize}

\item a real-valued implementation and specification language for numerical
programs with uncertainties; we define its semantics in terms of verification
constraints that they induce.

\item an approximation procedure for computing precise range bounds for nonlinear
expressions which combines SMT solving with interval arithmetic.

\item an approach for sound and fully \emph{automatic} error estimation for nonlinear
expressions for floating-point as well as fixed-point arithmetic of various precisions.
We handle roundoff and propagation errors separately with affine
arithmetic and a first-order Taylor approximation, respectively. While providing
accuracy, this separation also allows us to provide the programmer with useful
information about the numerical stability of the computation.

\item an extension of the error estimation to programs with simple loops,
where we developed a technique to express the total error as a function of the
number of loop iterations.

\item a sound and scalable technique to estimate discontinuity errors which
crucially relies on the use of a nonlinear SMT solver.

\item an open-source implementation in a tool called Rosa which we evaluate on
a number of benchmarks from the scientific computing and embedded systems
domain and compare to state-of-the-art tools.

\end{itemize}

%% file: overview.tex
\section{A Compiler for Reals}

We first introduce Rosa's specification language and give a high-level overview
of the technical challenges and our solutions on a number of examples.

\sparagraph{A Real-Valued Specification Language}
  Rosa is a `verifying' source-to-source compiler which takes as input a
  program written in a real-valued non-executable specification language and
  outputs appropriate finite-precision code. A program is written in a
  functional subset of the Scala programming language and consists of a number
  of methods over the \code{Real} data type.
  Figures~\ref{fig:sine},\ref{fig:pendulum} and \ref{fig:jet-approx} show
  three such example methods. Pre- and postconditions allow the user to
  explicitly state possible errors on method inputs and maximum tolerable
  errors on the output(s), respectively. Taking into account all uncertainties
  and their propagation, Rosa chooses a data type from a range of floating-point
  and fixed-point precisions and emits the corresponding implementation
  in the Scala programming language.

  By writing programs in a real-valued source language, programmers can reason
  about the correctness of the real-valued algorithm, and leave the low-level
  implementation details to the automated sound analysis in Rosa. Besides
  this separation of concerns, the real-valued semantics also serves as
  an unambiguous ideal baseline against which to compute errors. We believe
  that making numerical errors explicit in pre- and postcondition attached
  directly to the source code makes it less likely that these will be forgotten or
  overlooked. Finally, such a language opens the door to \emph{sound} compiler
  optimizations exploiting properties which are valid over reals, but not
  necessarily over finite-precision - as long as the accuracy specification is
  satisfied.

\sparagraph{Compilation Algorithm}
  If a full specification (pre- and postcondition) is present on a method,
  Rosa analyzes the numerical computation and selects a suitable finite-precision
  data type which fulfills this specification and synthesizes the
  corresponding code. The user can specify which data types are acceptable
  from a range of floating-point and fixed-point precisions. The order in
  which these possible data types are given is important. Rosa searches through
  the data types, applying a static analysis for each, and tries to
  find the first in the list which satisfies the specification.
  While this analysis is currently repeated for each data type, parts of
  the computation can be shared and we plan to optimize the compilation process
  in the future.

\begin{figure}
  \centering
    \includegraphics[width=0.8\textwidth]{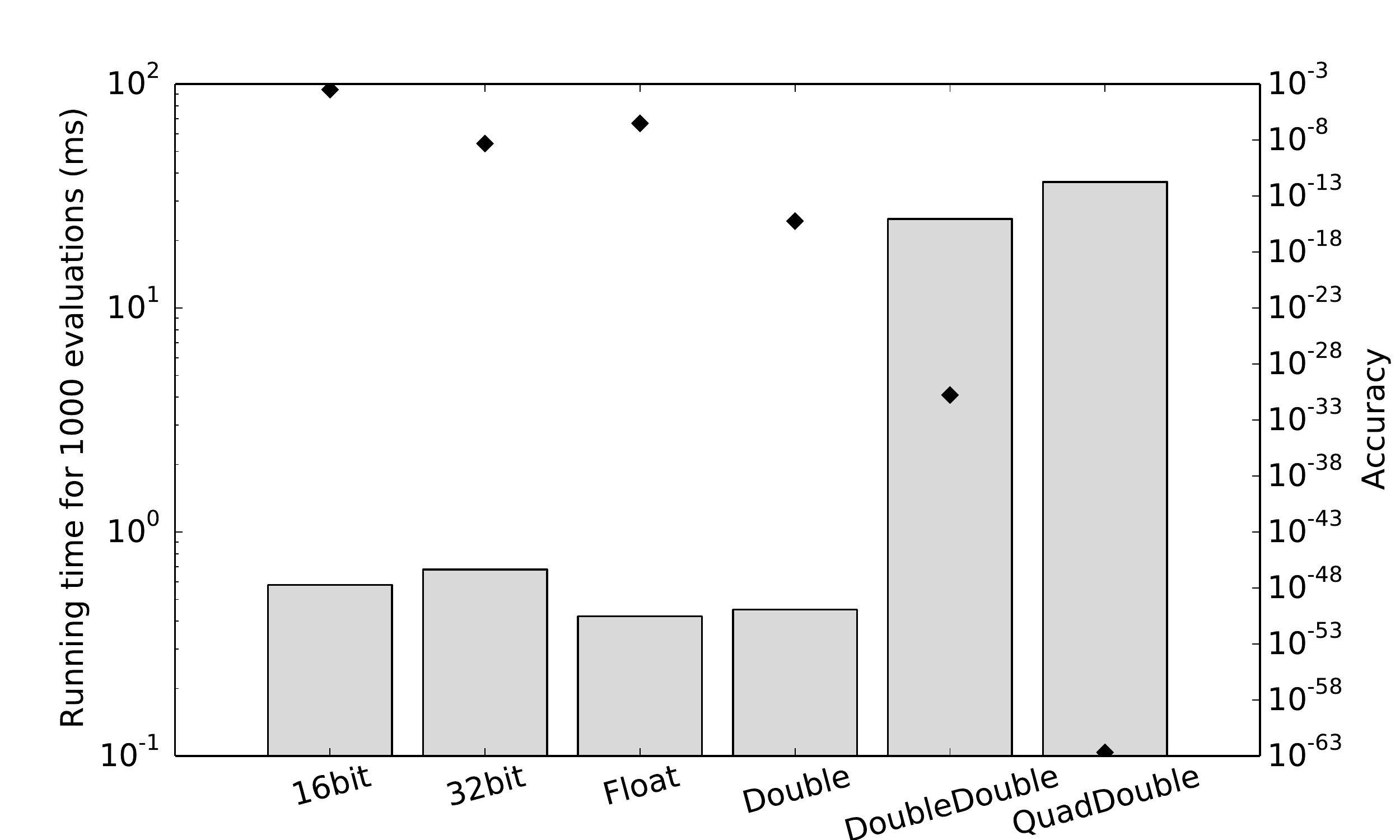}
    \caption{Running times vs accuracy for different finite-precision data types}
    \label{fig:tradeoff}
  \end{figure}

  The selection and the order of data types may depend on resource limits. If,
  for example, a floating-point unit is not available or would be very costly,
  the programmer can prioritize fixed-point data types. In general, the larger
  the data type (in terms of bits) the more time and memory the execution
  will take. We illustrate the performance vs. accuracy trade-off for the
  sine function from~\autoref{fig:sine} in the graph
  in~\autoref{fig:tradeoff}, which shows the runtime and accuracy and in
  particular the trade-off between the two for different data types. Rosa
  helps programmers navigate this trade-off space by providing automated
  sound accuracy estimates.

  Rosa can also be used as an analysis tool. By providing one data type
  (without necessarily a postcondition), Rosa will perform the analysis and
  report the results, consisting of a real-valued range and maximum error of
  the result, to the user. These analysis result are also reported during
  the regular compilation process as they may yield useful insights.

\begin{figure}
\begin{lstlisting}
def sineWithError(x: Real): Real = {
  require(x > -1.57079632679 && x < 1.57079632679 && x +/- 1e-11)

  x - (x*x*x)/6.0 + (x*x*x*x*x)/120.0 - (x*x*x*x*x*x*x)/5040.0
} ensuring(res => res +/- 1.001e-11)
\end{lstlisting}
  \caption{Approximation of sine with a Taylor expansion}
  \label{fig:sine}
  \end{figure}
  \begin{figure}
\begin{lstlisting}
/*
  @param x (x +/- 1.e-11)
  @return ((-1.0003675439019308 <= result && result <= 1.0003675439019308 &&
          (result +/- 3.8907801077969955e-09)))
*/
def sineWithError(x : Long): Long = {
  require(-1.5707963268 <= x && x <= 1.5707963268)

  val _tmp1 = ((x * x) >> 31)
  val _tmp2 = ((_tmp1 * x) >> 30)
  val _tmp3 = ((_tmp2 << 30) / 1610612736l)
  val _tmp4 = ((x << 1) - _tmp3)
  val _tmp5 = ((x * x) >> 31)
  val _tmp6 = ((_tmp5 * x) >> 30)
  val _tmp7 = ((_tmp6 * x) >> 31)
  val _tmp8 = ((_tmp7 * x) >> 31)
  val _tmp9 = ((_tmp8 << 28) / 2013265920l)
  val _tmp10 = ((_tmp4 + _tmp9) >> 1)
  val _tmp11 = ((x * x) >> 31)
  val _tmp12 = ((_tmp11 * x) >> 30)
  val _tmp13 = ((_tmp12 * x) >> 31)
  val _tmp14 = ((_tmp13 * x) >> 31)
  val _tmp15 = ((_tmp14 * x) >> 30)
  val _tmp16 = ((_tmp15 * x) >> 31)
  val _tmp17 = ((_tmp16 << 23) / 1321205760l)
  val _tmp18 = (((_tmp10 << 1) - _tmp17) >> 1)
  val result = _tmp18
}
\end{lstlisting}
    \caption{Sine function implemented in fixed-point arithmetic}
    \label{fig:sine-fixed}
  \end{figure}

\subsection{Example 1: Straight-line Nonlinear Computations}

  We illustrate this compilation process on the example in~\autoref{fig:sine},
  which shows the code of a method which computes the sine function
  with a 7th order Taylor approximation. The \code{Real} data type denotes an
  ideal real-valued variable without uncertainty or roundoff. The \code{require}
  clause specifies the range of the input parameter \code{x} as well as an
  initial uncertainty of 1e-11, which may stem from previous computations or
  measurement imprecisions. We would like to make sure that error on the result
  does not grow too large, so we constrain the error to 1.001e-11.
  We leave the data type unconstrained, so that Rosa considers 8, 16 and 32bit
  fixed-point arithmetic as well as single, double, double double and quad
  double floating-point precision by default (in this order). Rosa determines
  that 8 bit fixed-point precision potentially overflows, 16 or 32 bit fixed-point
  arithmetic is not accurate enough (total error of 2.54e-4 and 3.90e-9
  respectively) and that single-precision floating-point arithmetic is neither
  (error of 2.49e-7). For double floating-point precision, Rosa determines the
  error of 1.000047e-11, so that it generates code over the \code{Double} data
  type.
  If we could accept a somewhat larger error, say 5e-9, then Rosa determines
  that 32bit fixed-point arithmetic is sufficient. We show the generated code
  in~\autoref{fig:sine-fixed}. (All intermediate results fit into 32 bits,
  but as we need up to 64 bits to perform the arithmetic operations, we simulate
  the arithmetic here with the 64 bit \code{Long} data type.)
  Note that the generated code retains part of the precondition. In Scala,
  \code{require} expressions are actually runtime checks that throw
  an exception if the given condition is violated. Our error analysis, and code
  generation as well, is only valid if the range and error of the input is satisfied.
  We can only check the ranges at runtime, and Rosa generates the constraints
  such that they check the actual range (i.e. the ideal real-valued ranges $\pm$
  all possible errors). Checking the result range with a postcondition
  is not necessary, since Rosa has already proven those bounds correct.

  In order to determine which data type is appropriate, Rosa performs a static
  analysis which computes a sound estimate of the worst-case absolute error.
  Since roundoff errors and error propagation depend on the ranges of (all
  intermediate) computations, Rosa needs to compute these as accurately as
  possible. We developed a technique which combines interval
  arithmetic~\cite{Moore1966} with a nonlinear SMT solver, which provides
  accuracy as well as automation (\autoref{sec:ranges}). In addition, using an
  SMT solver allows Rosa to take into account arbitrary additional constraints
  on the inputs, which, for example, neither interval nor affine
  arithmetic~\cite{Figueiredo2004} can.
  Rosa decomposes the total error into roundoff errors and propagated initial
  errors and computes each differently (\autoref{sec:straightline}). The
  accumulation of roundoff errors is kept track off with affine arithmetic,
  while propagation errors are soundly estimated with a first-order Taylor
  approximation. The latter is fully automatically and accurately computed
  with again the help of a nonlinear SMT solver.

\subsection{Example 2: Loops with Constant Ranges}

  \begin{figure}
\begin{lstlisting}
def sine(x: Real): Real = {
  require(-20 <= x && x <= 20)
  x - x*x*x/6 + x*x*x*x*x/120
}

def pendulum(t: Real, w: Real, n: LoopCounter): (Real, Real) = {
  require(-2 <= t && t <= 2 && -5 <= w && w <= 5 &&
    -2.01 <= ~t && ~t <= 2.01 && -5.01 <= ~w && ~w <= 5.01)

  if (n < 100) {
    val h: Real = 0.01
    val L: Real = 2.0
    val m: Real = 1.5
    val g: Real = 9.80665

    val k1t = w
    val k1w = -g/L * sine(t)
    val k2t = w + h/2*k1w
    val k2w = -g/L * sine(t + h/2*k1t)
    val tNew = t + h*k2t
    val wNew = w + h*k2w

    pendulum5(tNew, wNew, n++)
  } else {
    (t, w)
  }
}
\end{lstlisting}
    \caption{Simulation of a pendulum}
    \label{fig:pendulum}
  \end{figure}

  In general, numerical errors in loops grow unboundedly and the state-of-the-art
  to compute sound error bounds in complex code is by unrolling. It turns
  out, however, that our separation of errors into roundoff and propagation
  errors allows us to express the error as a function of the number of loop
  iterations. We have identified a class of loops for which we can derive a
  closed-form expression of the loop error bounds. This expression, on one
  hand, constitutes an inductive invariant, and, on the other hand, can be
  used to compute concrete error bounds. While this approach is limited to
  loops where the variable ranges are bounded, our experiments show that this
  approach can already analyze interesting loops that are out of reach for
  current tools.

  ~\autoref{fig:pendulum} shows such an example: a Runge Kutta order 2
  simulation of a pendulum. \code{t} and \code{w} are the angle the pendulum
  forms with the vertical and the angular velocity respectively. We approximate
  the sine function with its order 5 Taylor series polynomial. We focus on
  \emph{roundoff} errors between the system following the real-valued dynamics
  and the system following the same dynamics but implemented in finite precision
  (we do not attempt to capture truncation errors due to the numerical
  integration, nor due to the Taylor approximation of sine). After 100 iterations,
  for instance, Rosa determines that the error on the results is at most 8.82e-14
  and 1.97e-13 (for \code{t} and \code{w} respectively)
  when implemented in double-precision floating-point arithmetic and
  7.38e-7 and 1.65e-6 in 32 bit fixed-point arithmetic.

\subsection{Example 3: Discontinuities}

  \begin{figure}
\begin{lstlisting}
def jetApproxGoodFitWithError(x: Real, y: Real): Real = {
 require(-5<=x && x<=5 && -5<=y && y<=5 && x +/- 0.001 && y +/- 0.001)
  if (y < x) {
    -0.317581 + 0.0563331*x + 0.0966019*x*x + 0.0132828*y +
      0.0372319*x*y + 0.00204579*y*y
  } else {
    -0.330458 + 0.0478931*x + 0.154893*x*x + 0.0185116*y -
      0.0153842*x*y - 0.00204579*y*y
  }
}
\end{lstlisting}
    \caption{Approximation of a complex embedded controller}
    \label{fig:jet-approx}
  \end{figure}

  Embedded systems often use piece-wise approximations of more complex
  functions. In~\autoref{fig:jet-approx} we show a possible piece-wise
  polynomial approximation of a fairly complex jet engine controller. We
  obtained this approximation by fitting a polynomial to a sample of values of
  the original function. Note that the resulting function is not continuous.

  A precise constraint encoding the difference between the real-valued and
  finite-precision computation, if they take different paths, features
  variables that are tightly correlated. This makes it hard for SMT-solvers
  to cope with and makes linear approaches imprecise. We explore the
  separation of errors idea in this scenario as well, to soundly estimate errors due to
  conditional branches. We separate the real-valued difference from
  finite-precision artifacts. The individual error components are easier to
  handle individually, yet preserve enough accuracy.

  In our example, the real-valued difference between the two branches is
  bounded by 0.0428 (making it arguably a reasonable approximation given the large possible range of
  the result). However, this is not a sound estimate for the discontinuity
  error in the presence of roundoff and initial errors (in our example
  0.001).  With Rosa, we can confirm that the discontinuity error is
  bounded by 0.0450 with double floating-point precision, with all errors taken into account.

%% file: problem_def.tex
\section{Problem Definition}

  Clearly, error computation is the main technical challenge of our compiler
  for \code{Real}s. Before we describe our solution in detail, we first
  precisely define the error computation problem that Rosa is solving
  internally. Formally, an input program consists of one or more methods given
  by the grammar in~\autoref{fig:grammar}.
  \begin{figure}[b]
  \begin{lstlisting}[language=pseudo,numbers=none]
    P $::=$ def mName(args): res = {
              require(A$_1$ $\land$ $\ldots$ $\land$ A$_n$)
              ( L | D | B )
            }
    A $::=$ C | $x$ +/- const | S
    S $::=$ S $\land$ S | S $\lor$ S | $\neg$ S | C
    L $::=$ if (n < const) mName(B, n + 1) else args
    D $::=$ if (C) D else D | B
    B $::=$ val $x$ = F; B | F
    F $::=$ F + F | F - F | F * F | F / F | $\sqrt{\text{F}}$ | X
    C $::=$ F $\le$ F | F < F | F $\ge$ F | F > F
    X $::=$ $x$ | const
  \end{lstlisting}
  \caption{Rosa's input language}
  \label{fig:grammar}
  \end{figure}
  \code{args} denotes possibly multiple arguments and \code{res} can be a tuple.
  The method body may consist of the standard arithmetic operators ${+, -, *,
  /, \sqrt{}}$, as well as immutable variable declarations (\code{val t1 = ...
  }), functional calls and conditionals. Note that square root is only
  supported for floating-point arithmetic. The specification language is
  functional, so we represent loops as recursive functions (denoted
  \lstinline|L|), where \code{n} denotes the integer loop iteration counter.
  For loop-free code \lstinline|D|, note that more complex conditions on branches
  can be expressed with nesting.

  Let us denote by $P$ a real-valued function representing our program and
  by $x$ its input. Denote by $\tl{P}$ the corresponding finite-precision
  implementation of the program, which has the same syntax tree but with
  operations interpreted in finite-precision arithmetic. Let $\tlx$ denote
  the input to this finite-precision program. The technical challenge of Rosa
  is to estimate the difference:
  \eqnnum{\label{eqn:p-constraint}}{
    \max_{x, \tlx} \;\;\norm{P(x) - \tl{P}(\tl{x})}
  }
  which denotes the absolute error of the result of the program. This
  error is crucial for selecting an implementation data type.

  The domains of $x$ and $\tlx$, over which this expression is to be
  evaluated, are given by the user-provided precondition in the \code{require}
  clause. It defines range bounds $x_i \in [a_i, b_i], \tlx_i \in [c_i, d_i]$
  for each component of the possibly multivariate input, as well as absolute
  error bounds on the inputs of the form \code{$x_i\;$ +/- $\;\lambda_i$} that
  define the relationship $\norm{ x - \tl{x} } \le \lambda$, understood
  component-wise.  If no errors are given explicitly, we assume roundoff as
  the initial error. The ability to specify initial errors in addition to
  roundoff is important for modular verification where the errors of one
  method may feed to following ones. Additionally, the \code{require} clause
  may specify further constraints on the inputs, such as \code{x*x + y*y <=
  20.0}.
  Method calls are handled either by inlining the postcondition or the whole
  method body.

  Corresponding to the syntactic program is a real-valued mathematical
  expression which is the input to our core error computation procedure.
  Concretely, the input consists of one or several real-valued functions $f:
  \R^m \to \R^n$ over some inputs $x_i \in \R$, representing the arithmetic
  expressions \lstinline|F|.  We denote by $f$ and $x$ the exact \emph{ideal}
  real-valued function and variables and by $\tlf: \R^m \to \R^n, \tlx_i \in
  \R$ their \emph{actual} finite-precision counter-parts. Note that for our
  analysis all variables are real-valued; the finite-precision variable
  $\tlx$ is considered as a noisy versions of $x$. We perform the error
  computation with respect to some fixed target precision in floating-point or
  fixed-point arithmetic; this choice gives error bounds for each individual arithmetic operation.

  When $P$ consists of a {\bf nonlinear arithmetic expression} alone
  (\code{F}), then~\autoref{eqn:p-constraint} reduces to bounding the absolute
  error on the result of evaluating $f(x)$ in finite precision arithmetic:
  $\max_{x, \tlx}\norm{f(x) - \tl{f}(\tl{x})}$ (\autoref{sec:straightline}).
  When the body of $P$ is a {\bf loop} (\code{L}), then the constraint reduces
  to computing the overall error after $k$-fold iteration $f^k$ of $f$, where
  $f$ corresponds to the loop body. We define for any function $H$:
  $H^0(x)=x$, $H^{k+1}(x)=H(H^k(x))$. We are then interested in bounding
  (\autoref{sec:loops}):
  \eqn{
    \max_{x,\tlx}\norm{ f^k(x) - \tlf^k(\tlx) }
  }
  For code containing branches (grammar rule \lstinline|D|),
  \autoref{eqn:p-constraint} accounts also for the {\bf discontinuity} error.
  For example, if we let $f_1$ and $f_2$ be the real-valued functions
  corresponding to the \code{if} and the \code{else} branch respectively with
  the \code{if} condition $c$, then, if $c(x) \land \lnot c(\tlx)$, the
  discontinuity error is given by $\norm{ f_1(x) - \tl{f}_2(\tl{x}) }$, i.e.,
  it accounts for the case where the real computation takes the if-branch, and
  the finite-precision one takes the else branch. The overall error on $P$
  from~\autoref{eqn:p-constraint} in this case must account for the maximum of
  discontinuity errors between all pairs of paths, as well as propagation and
  roundoff errors for each path (\autoref{sec:discontinuity}).

  \sparagraph{A note on relative error} Our technique soundly overestimates
  the absolute error of the computation. A sound estimate of the relative error
  can be computed from this and from the range of the result provided that the
  range does not include zero. Whenever this is the case, Rosa also reports the
  relative error in addition to the absolute one.

\subsection{Finite-precision Arithmetic}

  Rosa supports analysis and code generation for floating-point and fixed-point
  arithmetic of different precisions.
  We assume IEEE754 floating-point semantics. We regard overflow and underflow
  as errors and Rosa will report the possibility of these occurring as such.
  We assume rounding to nearest, however as long as the roundoff error can be
  determined from the range of possible values, our analysis can be straight-forwardly
  adapted. Code generation currently supports standard single and double
  floating-point arithmetic, as well as double-double and quad-double precisions
  implemented in software~\cite{QD2013}.

  Fixed-point arithmetic also represents a subset of the rationals, but unlike
  floating-point arithmetic does not require specialized hardware. Instead, it
  is implemented with integer operations only, which makes it attractive
  especially for resource-bound systems. The consequence of this is, however,
  that the alignment of (implicit) decimal points has to be performed manually
  with bit shift operations at compile time. For this, the \emph{global}
  ranges at each intermediate computation step have to be known, respectively
  have to be computed. For more details please see~\citeN{Anta2010}, whose
  fixed-point semantics we follow.

  While the input and output language is a subset of Scala, the analysis is
  programming language agnostic, providing the IEEE 754 standard is supported,
  and adapting Rosa to a different back-end would be straight-forward.

%% file: ranges.tex
\section{Computing Ranges Accurately}\label{sec:ranges}

The first step to accurately estimating roundoff and propagation errors is to
have a procedure to estimate ranges as tightly as possible. This is important
as these errors directly depend on the ranges of all, including intermediate,
values and coarse range estimates may result in inaccurate errors computed or
make the analysis impossible due to spurious potential runtime errors
(division by zero, etc.).

\subsection{Interval and Affine Arithmetic}

Traditionally, guaranteed computations have been performed with interval
arithmetic~\cite{Moore1966}. Interval arithmetic computes a bounding interval
for each basic operation as
\begin{equation*}
x \circ y = [ min(x \circ y), max( x \circ y)] \quad \quad  \circ \in \lbrace +, -, *, /  \rbrace
\end{equation*}
and analogously for square root. For longer computations, interval arithmetic
introduces over-approximations, as it cannot track correlations between variables
(e.g. $x - x \ne 0$).

Affine arithmetic~\cite{Figueiredo2004} partially addresses
this loss of correlation by representing possible values of variables as affine forms
\eqn{
  \hat{x} = x_0 + \sum_{i=1}^n x_i \epsilon_i
}
where $x_0$ denotes the \emph{central value} (of the represented interval) and
each \emph{noise symbol} $\epsilon_i$ is a formal variable denoting a deviation
from this central value, intended to range over $[-1, 1]$. The maximum magnitude
of each \emph{noise term} is given by the corresponding $x_i$.
The range represented by an affine form is computed as
\eqn{
  [\hat{x}] = [x_0 - rad(\hat{x}), x_0 + rad(\hat{x})] ,\quad \quad  rad(\hat{x}) = \sum_{i=1}^n |x_i|
}
Note that the sign of the $x_i$s does not matter in isolation, it does, however,
reflect the relative dependence between values. E.g., take $x = x_0 + x_1 \epsilon_1$, then
\eqn{
x - x &= x_0 + x_1 \epsilon_1 - (x_0 + x_1 \epsilon_1)
= x_0 - x_0 + x_1 \epsilon_1 - x_1 \epsilon_1 = 0
}
If we subtracted $x' = x_0 - x_1 \epsilon_1$ instead, the resulting interval
would have width $2 * x_1$ and not zero.
Linear operations are performed term wise and are computed exactly, whereas
nonlinear ones need to be approximated. Affine arithmetic can thus track
linear correlations, it is, however, not generally better than interval
arithmetic: e.g. $x*y$, where $x = [-5, 3], y = [-3, 1]$ gives $[-13, 15]$ in
affine arithmetic and $[-9, 15]$ in interval arithmetic.

\subsection{Range Estimation using Satisfiability Modulo Theories (SMT) Solvers}

  \begin{figure}
    \begin{lstlisting}[language=pseudo,numbers=none]
    def getRange(expr, precondition, precision, maxIterations):
      z3.assertConstraint(precondition)
      [aInit, bInit] = evalInterval(expr, precondition.ranges);

      //lower bound
      if z3.checkSat(expr < a + precision) == UNSAT
        a = aInit
        b = bInit
        numIterations = 0
        while (b-a) < precision $\wedge$ numIterations < maxIterations
          mid = a + (b - a) / 2
          numIterations++
          z3.checkSat(expr < mid) match
            case SAT $\Rightarrow$ b = mid
            case UNSAT $\Rightarrow$ a = mid
            case Unknown $\Rightarrow$ break
        aNew = a
      else
        aNew = aInit

      //upper bound symmetrically
      bNew = ...
      return: [aNew, bNew]
      \end{lstlisting}
    \caption{Algorithm for computing the range of an expression}
    \label{fig:range-alg}
  \end{figure}

  While interval and affine arithmetic are reasonably fast for range estimation,
  they tend to introduce over-approximations, especially if the input intervals
  are not sufficiently small. We improve over them by combining interval
  arithmetic with a nonlinear SMT (Satisfiability Modulo Theories) constraint solver 
  to obtain automation and accuracy.

  \autoref{fig:range-alg} shows the algorithm for computing the lower bound of a
  range. The computation for the upper bound is symmetric. For each range to be
  computed, our technique first computes an initial sound estimate of the range
  with interval arithmetic. It then performs an initial quick check to test
  whether the computed first approximation bounds are already tight. If not, it
  uses the first approximation as the starting point and then narrows down the
  lower and upper bounds using a binary search. At each step of the binary
  search our tool uses the nonlinear \textsf{nlsat} solver within Z3~\cite{De-Moura2008,Jovanovic2012}
  to confirm or reject the newly proposed bound.
  The search stops when either Z3 fails, i.e. returns unknown for a query or
  cannot answer within a given timeout, the difference between subsequent bounds
  is smaller than a precision threshold, or the maximum number of iterations is
  reached. This stopping criterion can be adjusted by the user.

  \sparagraph{Additional Constraints} In our approach, since we are using Z3 to
  check the soundness of bounds, we can assert the additional constraints and
  perform all checks with respect to all additional and initial constraints.
  This is especially useful when taking into account branch conditions from
  conditionals.

  \sparagraph{Optimizations} Calling an SMT solver is fairly expensive so we want
  to minimize the number of calls. The algorithm~\autoref{fig:range-alg}
  presents several direct knobs to do this: the maximum number of iterations and
  the precision of the range estimate. Through our experiments we have
  identified suitable default values, which seem to present a good trade-off
  between accuracy and performance. In addition to these two parameters, if we
  are only interested in the final range, we do not need to call Z3 and the
  algorithm in~\autoref{fig:range-alg} for every intermediate expression. In
  principle, we could call Z3 only on the full expression, however, we found
  that this resulted in suboptimal results as this expression very often was too
  complex. We found a good compromise in calling Z3 only every 10 arithmetic
  operations. All of these parameters can be adjusted by the user.

%% file: straightline.tex
\section{Soundly Estimating Numerical Errors in Nonlinear Expressions}\label{sec:straightline}

Now we address the first challenge of error estimation for a loop-free nonlinear
function without branches:
$\norm{f(x) - \tl{f}(\tl{x})} \text{ where } \norm{x - \tl{x}} \le \lambda, f: \R^m \to \R^n$
and where the ranges for $x$ and $\tlx$ are given by the precondition.

\subsection{Error Estimation with Affine Arithmetic}\label{sec:aa-errors}

  Consider first roundoff errors only that is, we are interested in $\norm{f(x) - \tl{f}(x)}$,
  where the input errors are zero. Our procedure executes the computation abstractly
  by computing an interval and an affine form for each AST node:
  \begin{lstlisting}
  (range: Interval, $\hat{err}$: AffineForm)
  \end{lstlisting}
  \code{range} represents the real-valued range, and $\hat{err}$ the accumulated
  worst-case errors, with essentially one noise term for each roundoff error
  (together with artifacts from nonlinear approximations). The actual finite-precision
  range is then given by \code{range} + $[\hat{err}]$, where
  $[\hat{err}]$ denotes the interval represented by the affine form.
  For each computation step, we compute the
  \begin{enumerate}
  \item new range with our range procedure from~\autoref{sec:ranges}
  \item propagation of already accumulated errors
  \item new roundoff error, which is then added to the propagated affine form.
  \end{enumerate}
  Since we compute the range at each intermediate node, we also
  check for possible overflows, underflows, division by zero or negative
  square root errors without extra effort.

  \sparagraph{Propagation of Errors with Affine Arithmetic}
  For linear operations, errors are propagated with the standard rules of
  affine arithmetic. For multiplication, division and square root the error
  propagation depends on the range of values, so that we have to adapt our
  computation to use the ranges computed with our Z3-backed procedure. In the
  following, we denote the real range of a variable $x$ by $[x]$ and its associated
  error by the affine form $\hat{err_x}$. When we write $[x] * \hat{err}_y$ we
  mean that the interval $[x]$ is converted into an affine form and the
  multiplication is performed in affine arithmetic.
  Multiplication is computed as
  \begin{align*}
    x * y &= ([x] + \hat{err_x})([y] + \hat{err_y})\\
    &= [x] * [y] + [x] * \hat{err_y} + [y] * \hat{err_x} +  \hat{err_x} * \hat{err_y} + \rho
  \end{align*}
  where $\rho$ is the new roundoff error.
  Thus the first term contributes to the ideal range and the remaining three to the error affine form.
  The larger the factors $[x]$ and $[y]$ are, the larger the finally computed errors will be
  so that a tight range estimation is important for accuracy.
  Division is computed as
  \begin{align*}
  \frac{x}{y} &= x* \frac{1}{y} = ([x] + \hat{err_x})([1/y] + \hat{err_{1/y}})\\
  &= [x] * [\frac{1}{y}] + [x] * \hat{err_{\frac{1}{y}}} + [\frac{1}{y}] * \hat{err_x} + \hat{err_x} * \hat{err_{\frac{1}{y}}} + \rho
  \end{align*}
  For square root, we first compute an affine approximation of square root as in
  our previous work~\cite{Darulova2011}:
  \eqn{
  \sqrt{x} = \alpha * x + \zeta + \theta
  }
  and then perform the affine multiplication term wise.

  \sparagraph{Roundoff Error Computation}
  Roundoff errors for floating-point arithmetic are computed at each computation step as
  \begin{lstlisting}
    $\rho$ = $\delta$ * maxAbs(totalRange)
  \end{lstlisting}
  where $\delta$ is the machine epsilon, and added to $\hat{err}$ as a fresh noise term.
  Note that this roundoff error computation makes our error computation parametric in
  the floating-point precision. Since we regard (and report) subnormal numbers as errors,
  this error abstraction is sound.
  For fixed-point arithmetic, roundoff errors are computed as
  \begin{lstlisting}
    $\rho$ = getFormat(totalRange, bitWidth).quantizationError
  \end{lstlisting}
  where the \code{getFormat} function returns the best fixed-point
  format~\cite{Anta2010} that can accommodate the range. This computation is
  parametric in the bit-width.

\subsection{Separation of Errors}\label{sec:separation}

  We could use the affine arithmetic based procedure to track all errors,not
  only roundoff errors, by simply adding the initial error as a fresh noise
  term at the beginning. Such an approach treats all errors equally: the
  initial errors are propagated in the same way as roundoff errors which are
  committed during the computation. We found, however, that the over-approximation
  introduced by affine arithmetic for nonlinear computations
  increases substantially as the magnitude of the noise terms (i.e. the
  errors) becomes larger.
  Instead, we separate the total error as follows:
  \eqnnum{\label{eqn:separation-straightline}}{\textstyle
    \begin{split}
   \norm{ f(x) - \tlf(\tlx) } &=
   \norm{ f(x) - f(\tlx) + f(\tlx) - \tlf(\tlx) }\\
   &\le \norm{ f(x) - f(\tlx) } + \norm{ f(\tlx) - \tlf(\tlx) }
   \end{split}
  }
  The first term, $\norm{ f(x) - f(\tlx) }$, captures the error on the result
  of $f$ caused by the initial error between $x$ and $\tlx$. The second term,
  $\norm{ f(\tlx) - \tlf(\tlx) }$, covers the roundoff error committed when
  evaluating $f$ in finite precision, but note that we can now compute this roundoff
  error on the same input $\tlx$. Thus, we separate the
  overall error into the propagation of existing errors, and the newly
  committed roundoff errors.
  We denote by $\sigma_f: \R^m \to \R^n$ the function which returns the roundoff
  error committed when evaluating an expression $f$ in finite-precision arithmetic:
  $\sigma_f(\tlx) = \norm{ f(\tlx) - \tlf(\tlx) }$. We omit the subscript
  $_f$, when it is clear from the context. Further, $g: \R^m \to \R^n$ denotes
  a function which bounds the difference in $f$, given a difference in its
  inputs: $\norm{ f(x) - f(y) } \le g(\norm{ x - y })$. When $m, n > 1$,
  the absolute values are component-wise, e.g. $g(\norm{x_1 - y
  _1}, \dots, \norm{x_m - y_m})$, but when it is clear from the context, we will
  write $g(\norm{ x - y })$ for clarity.
  Thus, the overall numerical error is given by:
  \eqnnum{\label{eqn:straightline-errors}}{
    \norm{ f(x) - \tlf(\tlx) } \le g(\norm{ x - \tlx }) + \sigma(\tlx)
  }
  One alternative to~\autoref{eqn:separation-straightline} would be to
  bound the error by $\norm{ f(x) - \tlf(x) }$ $+ \norm{
  \tlf(x) - \tl{f}(\tl{x}) }$. The first term now corresponds to roundoff
  errors, but the second requires bounding the difference of $\tlf$ over a
  certain input interval. In the separation that we have chosen, we need to
  compute the difference over the real-valued $f$. Note that $f$ is a simpler
  function than its finite-precision counterpart, and its analysis is reusable across different concrete implementations.

  The function $\sigma$ is instantiated with the affine arithmetic based
  procedure from~\autoref{sec:aa-errors}. Since roundoff errors are local, we
  found affine arithmetic suitable for this purpose. In contrast, the
  propagation of existing errors (function $g$) depends highly on the
  steepness of the function, so we want to capture as much global information,
  such as correlations between variables, as possible. This is only feasible
  when looking at the function as a whole.

\subsection{Propagation Errors}

  We instantiate~\autoref{eqn:straightline-errors} with $g(x) = K \cdot x$,
  i.e. $\norm{ f(x) - f(y) } \le K \norm{ x - y }$ which bounds the deviation
  on the result due to a difference in the input by a linear function in the input errors.
  The constant $K$ (or vector of constants $K_i$ in the case of a multivariate function)
  is to be determined for each function $f$ individually, and is
  usually called the Lipschitz constant. We will also use the in this
  context more descriptive name \emph{propagation coefficient}. Note that we
  need to compute the propagation coefficient $K$ for the mathematical
  function $f$ and not its finite-precision counterpart  $\tl{f}$.

  Error amplification or diminution depends on
  the derivative of the function at the \emph{value of the inputs}. The steeper the
  function, i.e. the larger the derivative, the more the errors are magnified.
  For $f: \R^m \to \R$ we have
  \eqnnum{\label{eqn:propagation-straightline}}{
    \norm{f(x) - f(\tlx)} \le \sum^m_{i=1} K_i \lambda_i, \qquad \text{ where }
    K_i = \sup_{x, \tlx}\nNorm{\partialDer{f}{w_i}}
  }
  where $\lambda_i$ are the initial errors and $w_i$ denote the formal parameters of $f$.
  This computation naturally extends component-wise to multiple outputs.
  Thus, the propagation coefficients are computed as a sound bound on the Jacobian.

  We formally derive the computation of the propagation coefficients $K_i$
  for a multivariate function $f: \R^m \to \R$ in the following.
  Let $h: [0, 1] \to \R$ such that $h(\theta) := f(y + \theta(z-y))$.
  Without loss of generality, assume $y < z$.
  Then $h(0) = f(y)$ and $h(1) = f(z)$ and
  $\frac{d}{d\theta}h(\theta) = \nabla f(y + \theta(z-y)) \cdot (z-y)$.
  By the mean value theorem:
  $f(z) - f(y) = h(1) - h(0) = h'(\zeta) \text{, where } \zeta \in [0, 1]$.

  \eqn{
     \norm{ &f(z) - f(y) } = \norm{ h'(\zeta) }
  = \norm{ \nabla f(y + \zeta(z-y)) \cdot (z-y) } \\
  &= \nNorm{ \left(\left.\partialDer{f}{w_1}\right|_s,  \dots,
   \left.\partialDer{f}{w_m}\right|_s \right) \cdot (z-y) },
   \quad s = y + \zeta(z-y)\\
  &= \nNorm{
  \partialDer{f}{w_1} \cdot (z_1 - y_1) + \dots + \partialDer{f}{w_m}  \cdot (z_m - y_m) }\\
  %
  &\le \sum^m_{i=1} \nNorm{\partialDer{f}{w_i} } \cdot \nNorm{z_i - y_i}  \qquad \text{(**)}
  }
  where the partial derivatives are evaluated at $s = y + \zeta(z-y)$ (which
  we omit for readability). The value of $s$ in (**) is constraint to be in $s
  \in [y, z]$, so for a sound analysis we have to determine the maximum
  absolute value of the partial derivative over $[y, z]$. $y$ and $z$ in our
  application range over the values of $x$ and $\tlx$ respectively, so we
  compute the maximum absolute value of $\partialDer{f}{x_i}$ over
  all possible values of $x$ and $\tlx$.
  With $\norm{y_i - z_i} \le \lambda_i$ we obtain
  \eqn{
    \norm{f(x) - f(\tlx)} \le \sum^m_{i=1} K_i \lambda_i, \qquad \text{ where }
    K_i = \sup_{x, \tlx}\nNorm{\partialDer{f}{w_i}}
  }

  \sparagraph{Bounding Partial Derivatives}
  We compute the partial derivatives symbolically. Recall that the arithmetic operations
  permitted are $\{+, -, *, /, \sqrt{}\}$, which leaves the possibility of discontinuities and undefined expressions. We detect these automatically during the bound computation, so we do not
  need to make or check any assumptions on the derivatives up-front.

  We need to soundly bound the partial derivatives over all possible values
  of $x$ and $\tlx$. Both interval and affine arithmetic suffer from
  possibly large over-approximations due to nonlinearity and loss of
  correlations. Furthermore, they cannot take additional constraints into account, for
  example from branch conditions (e.g. \code{y < x}) or user defined
  constraints on the inputs. We use the range computation from~\autoref{sec:ranges}
  which allows us to take these into account, making the ranges computed much tighter.

  \sparagraph{Sensitivity to Input Errors} Beyond providing a way to compute the
  propagated initial errors, ~\autoref{eqn:propagation-straightline} also makes
  an upper bound on the sensitivity of the function to input errors explicit.
  The user can use this knowledge, for example, to determine which inputs need
  to be determined more precisely, e.g. by more precise measurements or by using
  a larger number of iterations of a numerical algorithm to find them. We report
  the values of $K$ back to the user.

\subsection{Relationship with Affine Arithmetic}\label{sec:aaVsLipschitz}
  Both our presented propagation procedure and propagation using affine
  arithmetic perform approximations. The question arises then, when is it
  preferable to use one over the other? Our experience and experiments show
  empirically that for longer nonlinear computations, error propagation based
  on Lipschitz continuity gives better results, whereas for shorter and linear
  computations this is not the case. In this section, we present an analysis
  of this phenomenon based on a small example.

  Suppose we want to compute $x*y - x^2$. For this discussion we consider
  propagation only and disregard roundoff errors. We consider the case where
  $x$ and $y$ have an initial error of $\delta_x \epsilon_1$ and $\delta_y
  \epsilon_2$ respectively, where $\epsilon_i \in [-1, 1]$ are the formal
  noise symbols of affine arithmetic. Without loss of generality, we assume
  $\delta_x, \delta_y \ge 0$. We first derive the expression for the error
  with affine arithmetic and take the definition of multiplication
  from~\autoref{sec:aa-errors}. We denote by $[x]$ the evaluation of the \emph
  {real-valued} range of the variable $x$.

  The total range of $x$ is then the real-valued range plus the error: $[x] + \delta_x \epsilon_1$, where
  $\epsilon_1 \in [-1, 1]$. Multiplying out, and removing the $[x][y] - [x]^2$
  term (since it is not an error term), we obtain the expression for the error of $x*y - x^2$:

  \eqnnum{\label{eqn:affine-prop}}{
  \begin{split}
  \left([y] \delta_x \epsilon_1 + [x] \delta_y \epsilon_2 + \delta_x \delta_y \epsilon_3\right) -
  \left(2[x] \delta_x \epsilon_1 + \delta_x \delta_x \epsilon_4  \right) \\
  = ([y] - 2[x])\delta_x \epsilon_1 + [x]\delta_y \epsilon_2 + \delta_x\delta_y\epsilon_3 +
   \delta_x\delta_x\epsilon_4
   \end{split}
  }
  $\epsilon_3$ and $\epsilon_4$ are fresh noise symbols introduced by the nonlinear approximation.
  Now we compute the propagation coefficients:
  \eqn{
  \partialDer{f}{x} = y - 2x \quad \quad \partialDer{f}{y} = x
  }
  so that the error is given by
  \eqnnum{\label{eqn:lipschitz-prop}}{
   \Big|[y + \delta_y \epsilon_2 - 2 (x + \delta_x \epsilon_1) ]\Big| \delta_x +
    \Big|[x + \delta_x \epsilon_1]\Big| \delta_y
  }
  We obtain this expression by instantiating Equation (**) with the range expressions
  of $x$ and $y$.
  Note that the ranges used in the evaluation of the partial derivatives
  include the errors.
  Multiplying out~\autoref{eqn:lipschitz-prop} we obtain:
  \eqnnum{\label{eqn:prop-lipschitz-mult}}{
  \Big| [y - 2x] \Big| \delta_x + \Big|[x]\Big| \delta_x +  \delta_x \delta_y + \delta_x \delta_x + \delta_x \delta_x
  }
  With affine arithmetic, we
  compute ranges for propagation at each computation step, i.e. in~\autoref{eqn:affine-prop}
  we compute $[x]$ and $[y]$ separately.
  In contrast, with our new technique, the range is computed once, taking
  all correlations into account between the variables $x$ and $y$.
  It is these correlations that improve the computed error bounds.
  For instance, if we choose $x \in [1, 5]$ and $y \in [-1, 2]$ and
  we, say, know that $x < y$, then by a step-wise computation we obtain
  $[y] - 2[x] = [-1, 2] - 2[1, 5] = [-11, 0]$
  whereas taking the correlations into account, we can narrow down
  the range of $x$ to $[1, 2]$ and obtain
  $[y - 2x] = [-1, 2] - 2[1, 2] = [-5, 0]$.
  Hence, since we compute the maximum absolute value of these ranges for the error
  computation, affine arithmetic will use the factor 11, whereas our approach will use 5.

  But, comparing~\autoref{eqn:prop-lipschitz-mult} with~\autoref{eqn:affine-prop}, we also see that one term
  $\delta_x\delta_x$ is included twice with our approach, whereas in the affine propagation it
  is only included once.
  We conclude that a Lipschitz-based error propagation is most useful for
  longer computations where it can leverage correlations. In other cases, we
  keep the existing affine arithmetic-based technique. It does not require
  a two-step computation, so we want to use it for smaller expressions.
  We remark that for linear operations the two approaches are equivalent.

\subsection{Implementation}

  We have implemented Rosa in the Scala programming language. Internally, we
  use a rational data type implemented on top of Java's \code{BigInteger}s for
  all our computations. This lets us avoid having to deal with roundoff errors
  ourselves, and easily interface with Z3 which also uses rationals.

\subsection{Comparison with State-of-the-Art}

  We are aware of two other tools which can automatically quantify numerical
  errors: Fluctuat~\cite{Goubault2011} and FPTaylor~\cite{Solovyev2015}.

  Fluctuat is an abstract interpreter which uses affine arithmetic for both
  the ranges of variables and for the error computation. In order to combat
  the over-approximations introduced by affine arithmetic, Fluctuat can add
  constraints on noise terms~\cite{Ghorbal2010}. Further, Fluctuat uses Taylor
  approximations locally to handle division and square root~\cite{Ghorbal2009},
  but the expansion is hard coded and does not consider the global expression

  Another technique employed by Fluctuat is interval subdivision, where the
  user can designate up to two variables in the program whose ranges will be
  subdivided, analyzed separately and the results then merged. This procedure
  works for floating- point arithmetic as the decimal point is dynamic,
  however, for fixed-point arithmetic the global ranges are needed at each
  point. Naturally, interval subdivision increases the runtime of the
  analysis, especially for multivariate functions, and the optimal subdivision
  strategy may not always be obvious. Interval subdivision can also be
  straight-forwardly added to Rosa - at a performance penalty. We choose here
  to compare our SMT-based technique against Fluctuat with and without
  subdivision to obtain a good comparison between the techniques. In the
  future, we expect a combination of different techniques to work best.

  Fluctuat also has a procedure for computing discretization errors, and can
  handle loops either by computing fixpoints, if such exist, or by unrolling.
  Finally, Fluctuat also separates errors similarly to our presented approach,
  although it does not treat the different part fundamentally differently as
  we do. We want to note that our formalism has also enabled the unified
  treatment of loops and discontinuities.

  FPTaylor~\cite{Solovyev2015} is a recent tool for computing the roundoff
  errors of nonlinear expressions, including transcendental functions. It
  relies similarly to Rosa on Taylor series, but does the expansion with respect
  to errors, whereas we expand with respect to inputs. Furthermore,
  FPTaylor uses a global optimization as the backend, which enables the use of
  transcendental functions (Z3's nlsat solver only supports arithmetic).
  FPTaylor currently only supports error computation for  straight-line
  computations in floating-point arithmetic.

  Like Rosa, both Fluctuat and FPTaylor also compute relative errors from
  absolute errors, whenever the resulting range does not straddle zero.

  Another framework that can be used for estimating numerical errors is the
  Frama-C framework~\citeyear{framaC2015} with the Gappa front-end~\cite{Boldo2011}.
  Gappa works internally with interval arithmetic and
  works best on precise properties when the user can provide
  hints~\cite{Solovyev2015}. In this paper we want to focus on automated error
  estimation, thus we only compare our results against those from Fluctuat and
  FPTaylor.

\subsection{Experimental Results}

  \urlstyle{sf}

  We have chosen a number of benchmarks from the domains of scientific
  computing and embedded systems~\cite{Anta2010,Woodford2012} to evaluate the
  accuracy and performance of our technique. The tool and all benchmarks are
  open-source and available at \url{https://github.com/malyzajko/rosa}.

  We perform all tests in double floating-point precision as this is the
  precision supported by both Fluctuat and FPTaylor. Rosa is currently the
  only tool that also supports fixed-point arithmetic. In our experience,
  while the absolute errors naturally change with varying precisions and data
  types, relative differences when comparing different tools on the same
  precision data type remain similar. Experiments were performed on a desktop
  computer running Ubuntu 14.04.1 with a 3.5GHz i7 processor and 16GB of RAM,
  and using the unstable branch (as of 10 December 2014) of Z3.

  \input{straightline-bigtable}
  ~\autoref{tab:straightline-errors} shows our experimental results in terms of
  accuracy (absolute errors computed) and performance (running time of tool).
  All running times have been rounded up.
  We consider three flavors of our benchmarks: inputs with roundoff errors only,
  inputs with initial larger uncertainties and inputs with an additional nonlinear
  constraint.

  \sparagraph{Inputs with Roundoff} In the first set of benchmarks
  in~\autoref{tab:straightline-errors} we assume only roundoff as the initial
  error on inputs. We compare against Fluctuat without and with subdivisions.
  For the subdivisions, we uniformly chose 20 subdivisions for the two inputs
  where the effect was largest. While choosing more is certainly possible, we
  found the running time increasing rapidly and disproportionately with the
  accuracy gains. For FPTaylor we used default settings with the improved
  rounding model, approximate optimization and the branch and bound optimization
  procedure, which we believe are the most accurate settings.
  The annotation '(ref)' marks benchmarks that are refactored. Rosa's
  technique in general benefits from such a refactoring
  (see~\autoref{sec:aaVsLipschitz}), which is supported by the experiments.
  For Fluctuat this is also sometimes the case when subdivisions are used.

  FPTaylor is able to compute the tightest error bounds on these benchmarks,
  but we observe that the difference (except for the jet example) are in many
  cases not very large. FPTaylor's computation is also the most
  time consuming in the majority of cases.

  Finally, we would like to remark that subdivisions cannot be directly
  extended to fixed-point arithmetic as the determination of fixed-point
  formats and thus roundoff errors requires the knowledge of the global
  ranges, i.e. the ranges valid over all input and not only over the
  subdivision.

  \sparagraph{Inputs with Uncertainties}
  The second set of benchmarks features inputs with uniform uncertainty of 1e-11,
  aiming to compare the different tools ability to estimate the error propagation
  accurately. The tools' settings are the same as for the first set of testcases.
  Except for the jet example, which is difficult for Z3, Rosa computes essentially
  as tight error bounds as FPTaylor with a smaller running time.

  \sparagraph{Inputs with Nonlinear Constraint} For the last set of benchmarks,
  we have constrained the inputs with a nonlinear constraint of the form $x*x +
  y*y + z*z < c$, where $x, y, z$ are input variables and $c$ is a meaningful
  benchmark-specific constant. This constraint is representative of constraints
  that cannot be captured by a linear technique like affine arithmetic. In Rosa,
  this constraint can be specified naturally in the precondition. In Fluctuat,
  it is possible to enclose the computation in an if-condition (\code{if
  (constr) {...}}) and the affine terms will be constrained with a linearized
  branch condition. We used the 'Constraints on noise symbols' setting. FPTaylor
  provides syntax to specify additional constraints, however these are only
  supported with Z3 as the backend, and hence without the improved rounding
  model. We observe that no one tool consistently provides the most accurate
  error estimates, but that FPTaylor's technique turns out to be quite expensive
  in this case.

%% file: straightline-bigtable.tex
\begin{table}%
\renewcommand*{\arraystretch}{1.2}
\tbl{Absolute errors computed by Rosa, Fluctuat and FPTaylor
 for double-precision floating-point arithmetic. (r) marks refactored benchmarks,
 (e) marks benchmarks with additional input errors\label{tab:straightline-errors}}{%
\begin{tabular}{@{}p{1.9cm}llp{1.2cm}lp{0.6cm}p{1cm}p{1.1cm}l@{}}
\toprule
benchmark & Rosa  & Fluctuat  & Fluctuat (subdiv) & FPTaylor & Rosa & Fluctuat & Fluctuat (subdiv)& FPTaylor \\
\multicolumn{9}{l}{with roundoff errors only}\\
\midrule
doppler & 4.15e-13  & 3.90e-13  & 1.54e-13  & 1.35e-13  & 8 & 1 & 2 & 7 \\
doppler (r) & 2.42e-13  & 3.90e-13  & 1.40e-13  & 1.35e-13  & 7 & 1 & 2 & 7 \\
jet & 5.33e-9  & 4.08e-8  & 2.10e-11  & 1.17e-11  & 95  & 1 & 2 & 12  \\
jet (r) & 4.91e-9  & 4.08e-8  & 1.88e-11  & 1.17e-11  & 77  & 1 & 2 & 12  \\
rigidBody  & 3.65e-11  & 3.65e-11  & 3.65e-11  & 3.61e-11  & 1 & 1 & 2 & 6 \\
rigidBody (r)  & 3.65e-11  & 3.65e-11  & 3.65e-11  & 3.61e-11  & 1 & 1 & 2 & 5 \\
sine  & 5.74e-16  & 7.97e-16  & 7.41e-16  & 5.52e-16  & 2 & 1 & 1 & 6 \\
sineOrder3  & 9.96e-16  & 1.15e-15  & 1.09e-15  & 8.90e-16  & 1 & 1 & 1 & 4 \\
sqroot  & 2.87e-13  & 3.21e-13  & 3.21e-13  & 2.87e-13  & 1 & 1 & 1 & 7 \\
turbine1  & 5.99e-14  & 9.20e-14  & 2.21e-14  & 2.11e-14  & 5 & 1 & 2 & 8 \\
turbine1 (r)  & 5.15e-14  & 9.26e-14  & 2.21e-14  & 2.11e-14  & 2 & 1 & 2 & 8 \\
turbine2  & 7.68e-14  & 1.29e-13  & 2.87e-14  & 2.62e-14  & 2 & 1 & 2 & 6 \\
turbine2 (r)  & 6.30e-14  & 1.34e-13  & 2.87e-14  & 2.62e-14  & 1 & 1 & 2 & 7 \\
turbine3  & 4.62e-14  & 6.99e-14  & 1.34e-14  & 1.55e-14  & 4 & 1 & 2 & 7 \\
turbine3 (r)  & 4.02e-14  & 7.03e-14  & 1.32e-14  & 1.55e-14  & 2 & 1 & 2 & 7 \\
  & & & &  total &         209 & 15  & 27  & 109 \\
  & & & &   total (-jet)&  37  & 13  & 23  & 85  \\
\multicolumn{9}{l}{with input errors}\\
\midrule
doppler (re) & 1.83e-11  & 5.45e-11  & 2.21e-11  & 1.82e-11  & 13  & 1 & 2 & 7 \\
jet (re) & 3.36e-7  & 4.67e-4  & 1.37e-7  & 3.85e-8  & 76  & 1 & 2 & 13  \\
turbine1 (re) & 4.61e-10  & 1.82e-9  & 6.02e-10  & 4.61e-10  & 2 & 1 & 2 & 8 \\
turbine2 (re) & 5.87e-10  & 2.82e-9  & 6.14e-10  & 5.86e-10  & 1 & 1 & 2 & 10  \\
turbine3 (re) & 3.33e-10  & 1.24e-9  & 2.53e-10  & 3.32e-10  & 2 & 1 & 2 & 8 \\
rigidBody (re)  & 1.50e-7  & 1.50e-7  & 1.50e-7  & 1.50e-7  & 2 & 1 & 2 & 6 \\
sine (e) & 1.00e-11  & 2.09e-11  & 1.01e-11  & 1.00e-11  & 2 & 1 & 1 & 6 \\
  &   &   &   & total  & 98  & 7 & 13  & 58  \\
  &   &   &   & total (-jet)  & 22  & 6 & 11  & 45  \\
\multicolumn{9}{l}{with input constraint}\\
\midrule
doppler (r)  & 1.76e-14  & 1.09e-13  & 4.84e-14  & 1.57e-14  & 7 & 1 & 2 & 5 \\
doppler (re) & 4.67e-13  & 1.37e-11  & 6.28e-12  & 4.77e-13  & 12  & 1 & 2 & 11  \\
jet (r)  & 4.91e-9  & 4.08e-8  & 1.88e-11  & 1.48e-11  & 84  & 1 & 2 & 1731  \\
jet (re) & 3.36e-7  & 4.67e-4  & 1.37e-7  & - & 81  & 1 & 2 & - \\
rigidBody (r) & 1.66e-11  & 3.65e-11  & 3.34e-11  & 1.52e-11  & 13  & 1 & 2 & 61  \\
rigidBody (re)  & 8.84e-8  & 1.50e-7  & 1.15e-7  & 6.78e-8  & 13  & 1 & 2 & 284 \\
turbine1 (r) & 4.26e-14  & 8.66e-14  & 2.21e-14  & 2.48e-14  & 3 & 1 & 2 & 6 \\
turbine1 (re)  & 4.61e-10  & 1.94e-9  & 6.51e-10  & 4.59e-10  & 2 & 1 & 2 & 109 \\
turbine2 (r) & 5.26e-14  & 1.45e-13  & 2.44e-14  & 2.92e-14  & 2 & 1 & 2 & 6 \\
turbine2 (re)  & 5.87e-10  & 3.02e-9  & 6.33e-10  & 4.84e-10  & 2 & 1 & 2 & 37  \\
turbine3 (r) & 3.55e-14  & 7.32e-14  & 9.50e-15  & 1.49e-14  & 5 & 1 & 2 & 11  \\
turbine3 (re)  & 3.33e-10  & 1.33e-09  & 2.30e-10  & 2.76e-10  & 5 & 1 & 2 & 316 \\
  &   &   &   & total & 229 & 12  & 24  & 2577  \\
  &   &   &   & total (-jet)  & 64  & 10  & 20  & 846 \\
\bottomrule
\end{tabular}}
\end{table}

%% file: loops.tex
\section{Loops}\label{sec:loops}

We have identified a class of loops for which the propagation of errors idea
allows us to express the numerical errors as a function of the number of
iterations. Concretely, we assume a single non-nested loop without conditional
branches for which the ranges of variables are bounded and fixed statically.
We do not attempt to prove that ranges are preserved across loop iterations;
we leave the discovery of suitable inductive invariants that implies ranges
for future work. Our approach does not include all loops, but it does cover a
number of interesting patterns, including simulations of initial value
problems in physics. We note that the alternative for analyzing numerical
errors in general nonlinear loops is unrolling, which, as our experiments
show, does not scale well.

\subsection{General Error Propagation}

  Representing the computation of the loop body by $f$, we want to compute the
  overall error after $k$-fold iteration $f^k$ of $f$: $\norm{ f^k(x) -
  \tlf^k(\tlx) }$. $f, g$ and $\sigma$ are now vector-valued: $f,
  g, \sigma: \R^n \to \R^n$, because we are nesting the potentially
  multivariate function $f$. In essence, we want to compute the effect of
  iterating~\autoref{eqn:straightline-errors}.\\

  {\bf Theorem: }
  Let $g$ be such that $\norm{ f(x) - f(y) } \le g(\norm{ x - y })$, it satisfies $g(x + y) \le g(x) + g(y)$
  and is monotonic. Further, $\sigma$ and $\lambda$ satisfy
  $\sigma(\tlx) \le \norm{ f(\tlx) - \tlf(\tlx) }$ and $\norm{x- \tlx} \le \lambda$. The
  absolute value is taken component-wise. Then the numerical error after $k$ iterations
  is given by
  \eqnnum{\label{eqn:propagation-loops}}{
   \norm{ f^k(x) - \tlf^k(\tlx) }  \le
    g^k(\norm{x - \tlx}) + \sum^{k - 1}_{i = 0} g^i(\sigma(\tlf^{k-i-1}(\tlx)))
  }
  Thus, the overall error after $k$ iterations can be decomposed into the initial error propagated
  through $k$ iterations, and the roundoff error from the $i^{th}$
  iteration propagated through the remaining iterations.

  {\bf Proof: }
  We show this by induction.  The base case $k = 1$ is covered by our treatment of
  straight-line computations (\autoref{sec:separation}).
  By adding and subtracting $f(\tlf^{k-1}(\tlx))_1$ we get
  \eqn{
     &\colvec{\norm{f^k(x)_1 - \tlf^k(\tlx)_1}}{\vdots}{\norm{f^k(x)_n - \tlf^k(\tlx)_n}} \\
    &\le
    \colvec{
      \norm{f^k(x)_1 - f(\tlf^{k-1}(\tlx))_1}
    }{\vdots}{
      \norm{f^k(x)_n - f(\tlf^{k-1}(\tlx))_n}
    }
     +
    \colvec{
      \norm{f(\tlf^{k-1}(\tlx))_1 -\tlf^k(\tlx)_1 }
    }{\vdots}{
    \norm{f(\tlf^{k-1}(\tlx))_n -\tlf^k(\tlx)_n }
    }
  }
  Applying the definitions of $g$ and $\sigma$
   \eqn{
    &\le g\colvec{
    \norm{f^{k-1}(x)_1 - \tlf^{k-1}(\tlx)_1}}{\vdots}{\norm{f^{k-1}(x)_n - \tlf^{k-1}(\tlx)_n}} +
    \sigma(\tlf^{k-1}(\tlx))
  }
  then using the induction hypothesis and monotonicity of $g$,
  \eqn{
    &\le g \left(g^{k-1}(\vec{\lambda}) + \sum^{k-2}_{i = 0} g^i(\sigma(\tlf^{k-i-1}(\tlx))) \right)+ \sigma(\tlf^{k-1}(\tlx))
  }
  then using $g(x+y) \le g(x) + g(y)$, we finally have
  \eqn{
    &\le g^k(\vec{\lambda}) + \sum^{k-1}_{i = 1} g^i(\sigma(\tlf^{k-i-1}(\tlx))) +
      \sigma(\tlf^{k-1}(\tlx))  \mspace{250.0mu} \\
    &= g^k(\vec{\lambda}) + \sum^{k - 1}_{i = 0} g^i(\sigma(\tlf^{k-i-1}(\tlx))) \mspace{180mu} \blacksquare
  }

  \subsection{Closed Form Expression}

    We instantiate the propagation function $g$ as before using propagation coefficients.
    Evaluating~\autoref{eqn:propagation-loops} as given, with a fresh set of
    propagation coefficients for each iteration $i$ amounts to loop unrolling,
    but with a loss of correlation between each loop iteration.
    We observe that when the ranges are bounded (as by our assumption), then we
    can  compute $K$ as a matrix of propagation coefficients, and similarly
    obtain $\sigma(\tlf^i)=\sigma$ as a vector of constants, both \emph{valid
    for all iterations}. Then we obtain a closed-form for the expression of the
    error:
    \eqn{
      \norm{ f^k(x) - \tl{f}^k(\tl{x})} &\le
         K^k \lambda + \sum^{k-1}_{i=1}K^i \sigma + \sigma
        = K^k \lambda + \sum^{k-1}_{i=0}K^i \sigma
    }
    where $\lambda$ is the vector of initial errors. Denoting by $I$ the identity matrix,
    if $(I-K)^{-1}$ exists,
    \vspace{-0.5em}
    \eqn{
      \norm{ f^k(x) - \tl{f}^k(\tl{x})}  &\le K^k \lambda + ( (I-K)^{-1}(I - K^k) )\sigma
    }
    We obtain $K^k$ with
    power-by-squaring and compute the inverse with the Gauss-Jordan method with
    rational coefficients to obtain sound results (though a closed-form is
    not strictly necessary for our purpose because we do know the number of iterations $k$).

    \sparagraph{Computing $K$ and $\sigma$}
    When the ranges of the variables of the loop are inductive, that is, both
    the real-valued and the finite-precision values remain within the initial
    ranges, then these are clearly the ranges for the computation of $K$ and
    roundoffs $\sigma$. For loops, we require the user to specify both the real-valued
    ranges of variables (e.g. \code{a <= x && x <= b}) as well as the actual
    finite-precision ones (\code{c <= ~x && ~x <= d}, as in~\autoref{fig:pendulum}).
    We also require that the actual ranges always
    include the real ones ($[a, b] \subseteq [c, d]$), and we use the actual
    ranges ($[c, d]$) for the computation of $K$ and $\sigma$. We believe that
    it is reasonable to assume that a user writing these applications to have
    the domain knowledge to be able to provide these specifications.

  \subsection{Handling Additional Sources of Errors}

    What if roundoff errors are not the only errors present? If the real-valued
    computation given by the specification is also the \emph{ideal} computation,
    we can simply add the errors in the same way as roundoff errors.  If the
    real-valued computation is, however, already an approximation of some other
    \emph{unknown} ideal function, say $f_*$, it is not directly clear how our error
    computation applies.

    This may be the case, for example, for truncation errors
    due to a numerical algorithm.
    To model such errors, let us suppose that we can compute (or at least overestimate) these by
    a function $\tau: \R^n \to \R^n$, i.e.
    $\tau_{\fs}(x) = |\fs(x) - f(x)|$.

    In the following we consider the one-dimensional case $n=1$ for simplicity of
    exposition, but it generalizes as before to the $n$-dimensional case.
    We can apply a similar separation of errors as before:
    \eqn{
    & \norm{\fs(x) - \tlf(\tlx) } \\
     & \qquad \le \norm{\fs(x) - f(x)} + \norm{f(x) - f(\tlx)} + \norm{f(\tlx) - \tlf(\tlx)}\\
     & \qquad = \tau(x) + g(\norm{x - \tlx}) + \sigma(\tlx)
    }
    which lets us decompose the overall error into the truncation error, the propagated
    initial error and the roundoff error.
    If we now iterate, we find by a similar argument as before:
    \eqn{
      & \norm{\fs^m(x) - \tlf(\tlx)} \\
      & \le g^m(\norm{x - \tlx}) + \sum^{m-1}_{j=0} g^j\left(\tau(\fs^{m-j-1}(x))\right) + g^j\left(\sigma(\tlf^{m-j-1}(\tlx))\right) \\
      & = g^m(\norm{x - \tlx}) + \sum^{m-1}_{j=0} g^j\left(\tau(\fs^{m-j-1}(x)) + \sigma(\tlf^{m-j-1}(\tlx))\right)
    }

    The result essentially means that our previously defined method can also be
    applied to the case when truncation (or similar) errors are present. We do
    not pursue this direction further however, and leave a proper automated
    treatment of truncation errors to future work.

\subsection{Experimental Results}

  We evaluate our technique on three benchmarks in~\autoref{tbl:loops}. We
  already presented the pendulum benchmark in~\autoref{fig:pendulum}. The mean
  benchmark computes a running average of values in a range of [-1200, 1200].
  The nbody benchmark is a 2-body simulation of Jupiter orbiting around the
  Sun. For each benchmark we consider different number of iterations of the
  loop and report the error for one of the loop's variables. Fluctuat is not
  able to compute a fixpoint for these benchmarks, as the errors keep growing
  with each iteration. Instead we manually set the number of times the loop is
  unrolled. For the pendulum 50 benchmark Rosa is able to compute a tighter
  error bound with a faster runtime. For larger numbers of iterations,
  Fluctuat reports an error of $\infty$. This is also the result for the nbody
  benchmark. For the mean benchmark, where the computation is less complex,
  Fluctuat can compute tighter error bounds, at the expense of much longer
  analysis times. This illustrates that our technique outperforms unrolling in
  Fluctuat for benchmarks that are highly nonlinear, whereas Fluctuat's
  strategy may be used for cases where the nonlinearity is limited as is the
  number of iterations. Note that Rosa's runtime is independent of the loop's
  number of iterations.

  \begin{table}%
  \renewcommand*{\arraystretch}{1.2}
  \tbl{Absolute errors and runtimes for different benchmarks
  and different number of loop iterations\label{tbl:loops}}{%
  \begin{tabular}{@{}lllcc@{}}
  \toprule
   & \multicolumn{2}{c}{Absolute errors} & \multicolumn{2}{c}{Running times} \\
  benchmark & Rosa  & Fluctuat  & Rosa  & Fluctuat  \\
  \midrule
  pendulum 50 & 2.21e-14  & 2.43e-13  & 8 & 47  \\
  pendulum 100  & 8.82e-14  & -  & 8 &  - \\
  pendulum 250  & 2.67e-12  & -  & 8 &  - \\
  pendulum 500  & 6.54e-10  & -  & 8 &  - \\
  pendulum 1000 & 3.89e-5  & -  & 8 &  - \\
  mean 100  & 3.21e-7  & 9.92e-9  & 5 & 1 \\
  mean 500  & 1.62e-6  & 1.01e-8  & 6 & 5 \\
  mean 1000 & 3.30e-6  & 1.01e-8  & 7 & 27  \\
  mean 2000 & 4.51e-6  & 1.03e-8  & 4 & 158 \\
  mean 3000 & 4.96e-6  & 1.05e-8  & 4 & 392 \\
  mean 4000 & 5.12e-6  & 1.06e-8  & 5 & 734 \\
  nbody 50  & 1.30e-11  & -  & 794 & -  \\
  nbody 100 & 1.35e-8  & -  & 776 &  - \\
  \bottomrule
  \end{tabular}}
  \end{table}%

%% file: discontinuities.tex
\section{Discontinuities}\label{sec:discontinuity}

  Recall the piece-wise jet engine approximation from~\autoref{fig:jet-approx}.
  Due to the initial errors on \code{x} and \code{y}, the real-valued
  computation may take a different branch than the finite-precision one, and
  thus produce a different result. We call this difference the
  \emph{discontinuity error}.

  We will assume that individual branch conditions are of the form \code{e1 $\circ$ e2}, where
  $\circ \in \{<, \le, >, \ge\}$ and \code{e1}, \code{e2} are arithmetic
  expressions. More complex conditions can be obtained by nesting conditionals.
  We do not assume the function represented by the conditional to be neither
  smooth nor continuous. We perform our analysis pairwise for each pair of
  paths in the program. While this gives, in the worst-case, an exponential
  number of cases to consider, we found that many of these paths are
  infeasible due to inconsistent branch conditions; such infeasible paths are eliminated
  early.

  \subsection{Applying Separation of Errors}

    Using our previous notation, let us consider a function with a single branch
    statement like in the example above and let $f_1$ and $f_2$ be the real-valued
    functions corresponding to the \code{if} and the \code{else} branch
    respectively. Then, the discontinuity error is given by $\norm{ f_1(x) -
    \tl{f}_2(\tl{x}) }$, i.e. the real computation takes branch $f_1$, and the
    finite-precision one $f_2$. The opposite case is analogous. We again apply
    the idea of separation of errors:
    \vspace{-0.5em}
    \eqnnum{\label{eqn:separation-discontinuity}}{
         \norm{ f_1(x) - \tlf_2(\tlx) }
         \le \norm{ f_1(x) - f_1(\tlx) } +
         \norm{ f_1(\tlx) - f_2(\tlx) } +
        \norm{ f_2(\tlx) - \tl{f}_2(\tlx) }
    }
    The individual components are
    \begin{enumerate}[topsep=0pt]
    \item
      $\norm{ f_1(x) - f_1(\tlx) }$: the difference in $f_1$ due to initial errors.
      We can compute this difference with our
      propagation coefficients: $\norm{ f_1(x) - f_1(\tlx) } \le K \norm{x - \tlx}$.

    \item
      $\norm{ f_1(\tlx) - f_2(\tlx) }$: the real-valued difference between $f_1$ and
      $f_2$. We can bound this value by the Z3-aided range computation
      from~\autoref{sec:ranges}.

    \item
      $\norm{ f_2(\tlx) - \tlf_2(\tlx) }$: the roundoff error when evaluating $f_2$
        in finite-precision arithmetic. We use the procedure from~\autoref{sec:aa-errors} as before.
    \end{enumerate}

    We expect the individual parts to be easier to handle for the underlying
    SMT-solver since we reduce the number of variables and correlations. We
    clearly introduce an additional over-approximation, but we observed in our
    experiments that this is in general small. In contrast, Fluctuat's
    approach relies on constraints on the affine forms to capture the
    different branch conditions~\cite{Goubault2013}. A split of the total
    error into two parts is also possible, e.g. as $\norm{ f_1(x) -
    \tlf_2(\tlx) } \le \norm{ f_1(x) - f_2(\tlx)} + \norm{ f_2(\tlx) -
    \tl{f}_2(\tlx)}$, which performs one computation less. This split,
    combined with a precise constraint relating $x$ to $\tlx$ would introduce
    \emph{one} constraint with many correlations between
    variables~\cite{Darulova2014}. Such a precise and complex relation
    overwhelms the SMT solver quickly, but bounding the ranges without the
    correlation information yields unsatisfactory results.

    \subsection{Determining Ranges for $x$ and $\tlx$}

      As in the previous sections, it is crucial to determine the ranges of
      $x, \tlx \in \R$ over which to evaluate the individual parts
      of~\autoref{eqn:separation-discontinuity}. A sound approach would be to
      simply use the input ranges, but this would lead to unnecessary over-approximations.
      In general, not all inputs can exhibit a divergence
      between the real-valued and the finite-precision computation. They are
      determined by the branch conditions and the errors on the variables.
      Consider the branch condition \code{if (e1 < e2)} and the case where the
      real-valued path takes the if-branch, i.e. variable $x$ satisfies $e1 <
      e2$ and $\tlx$ satisfies $e1 \ge e2$. The constraint for the finite-
      precision variables $\tlx$ is then

      \eqn{
        e1 + \delta_1 < e2 + \delta_2 \land e1 \ge e2
      }
      where $\delta_1, \delta_2$ are error intervals on evaluating $e1$ and $e2$ respectively.
      This constraint expresses that we want those values which satisfy the condition $e1 \ge e2$,
      but are ``close enough'' to the boundary such that their corresponding ideal real value could take
      the other path.
      We create such a constraint both for the variables representing finite-precision values ($\tlx$),
      as well as the real-valued ones $x$ and use them as additional constraints when computing the
      individual parts of~\autoref{eqn:separation-discontinuity}.
      The procedure for other branch conditions is analogous.

\subsection{Experimental Results}

  We evaluate our technique on a number of benchmarks with discontinuities,
  which we have either constructed by piece-wise approximating a more complex
  function or chosen from~\citeN{Goubault2013}. All the benchmarks' source
  code is available online. We compare our results in terms of accuracy and
  performance against Fluctuat. Fluctuat does not check for discontinuity
  errors by default; we enable this analysis with the 'Unstable test analysis'
  option (this is the only way). Subdivisions, however, do not appear to
  work with this setting.~\autoref{tbl:discontinuities} summarizes our
  results. While Fluctuat is faster than Rosa, Rosa is able to compute
  significantly tighter error bounds and, we believe, achieves a good
  compromise between accuracy and performance.

  \begin{table}%
  \renewcommand*{\arraystretch}{1.2}
  \tbl{Absolute discontinuity errors computed and runtimes of Rosa and Fluctuat\label{tbl:discontinuities}}{%
  \begin{tabular}{@{}lllcc@{}}
  \toprule
  benchmark & Rosa  & Fluctuat  & Rosa  & Fluctuat  \\
  \midrule
  cubicSpline & 1.25e-15  & 12.00 & 9 & 1 \\
  jetApprox & 0.0232  & 18.40 & 46  & 1 \\
  jetApprox (err) & 0.0242  & 19.06 & 44  & 1 \\
  jetApproxBadFit & 0.8825  & 9.305 & 16  & 1 \\
  jetApproxBadFit (err) & 0.8852  & 10.09 & 11  & 1 \\
  jetApproxGoodFit  & 0.0428  & 5.191 & 5 & 1 \\
  jetApproxGoodFit (err)  & 0.0450  & 5.193 & 5 & 1 \\
  linearFit & 0.6374  & 1.721 & 3 & 1 \\
  quadraticFit  & 0.2548  & 10.60 & 21  & 1 \\
  quadraticFit (err)  & 0.2551  & 10.96 & 20  & 1 \\
  quadraticFit2 & 3.14e-9  & 0.6321  & 4 & 1 \\
  quadraticFit2 (err) & 0.0009  & 0.7188  & 4 & 1 \\
  simpleInterpolator  & 3.40e-5  & 1.0e-5  & 1 & 1 \\
  sortOfStyblinski  & 1.0878  & 27.07 & 5 & 1 \\
  sortOfStyblinski (err)  & 1.0982  & 28.82 & 5 & 1 \\
  squareRoot  & 0.0238  & 0.0394  & 3 & 1 \\
  squareRoot3 & 2.76e-9  & 0.4289  & 6 & 1 \\
  squareRoot3Invalid  & 3.93e-9  & 0.4288  & 6 & 1 \\
  styblinski  & 4.81e-8  & 121.16  & 30  & 1 \\
  styblinski (err)  & 0.0132  & 124.10  & 25  & 1 \\
  \bottomrule
  \end{tabular}}
  \end{table}%

%% file: related.tex
\section{Related Work}

To the best of our knowledge, Fluctuat~\cite{Goubault2011,Goubault2013} and
FPTaylor~\cite{Solovyev2015} are most related to our work. We are not aware of
other tools or techniques that can \emph{soundly} and \emph{automatically}
quantify numerical errors in the presence of nonlinearity, branches and
loops.

In the context of abstract interpretation, domains exist that are sound with
respect to floating-points and that can be used to prove the absence of
runtime errors such as division by
zero~\cite{Blanchet2003,Mine2004,Feret2004,Chen2008,Ghorbal2009}.~\citeN{Feret2005}
presents an abstract domain which associates the ranges with the iteration
count, similar to our proposed technique for loops.~\citeN{Martel2002}
considers the stability of loops, by proving whether loops can asymptotically
diverge. The problem that we are solving is different, however, as we want to
quantify the \emph{difference} between the real-valued and the
finite-precision computation.

Floating-points have been formalized in the SMT-LIB format~\cite{Rummer2010},
and approaches exist which deal with the prohibiting complexity of bit-precise
techniques via approximations~\cite{Brillout2009,Haller2012}. Efficient combination of
theories needed to express roundoff errors is non-trivial, and we are not aware of an approach that is able to
quantify the deviation of finite-precision computations with respect to reals.
Floating-point precision assertions can also be proven using an interactive
theorem prover~\cite{Boldo2011,Linderman2010,Ayad2010,Harrison2006}. These
tools can reason about ranges and errors of finite-precision implementations,
but target specialized and precise properties, which, in general, require an
expert user and interactively guiding the proof. Very tight error bounds have
been shown by manual proof for certain special computations, such as
powers~\cite{Graillat2014}. Our work is on the other side of the trade-off
between accuracy and automation as well as generality.

Synthesis of specifically fixed-point arithmetic programs has also been an
area of active research, with different utilized techniques: simulation or
testing~\cite{Mallik2007,Jha2013}, interval or affine
arithmetic~\cite{Lee2006} or automatic differentiation~\cite{Gaffar2004}. Some
approaches try to optimize the bit-width whereas in our case we keep it fixed,
but provide a sound and accurate analysis, which could be used in combination
with an optimization technique, like e.g.~\citeN{Jha2013}. A similar approach
to our range estimation has been developed independently by~\citeN{Kinsman2009} in
the context of fixed-point arithmetic. We also identify the potential of
additional constraints and develop optimizations to make the use of an SMT
solver efficient enough. Further, our techniques aim to be generally
applicable to various finite-precision arithmetics.

Several approaches also exist to test the stability of numerical programs,
e.g. by perturbation of low-order bits and rewriting~\cite{Tang2010}, or by
perturbing the rounding modes~\cite{Scott2007}. Another common theme is to run
a higher-precision program alongside the original one.~\citeN{Benz2012} does
so by instrumentation,~\citeN{Paganelli2013} generates constraints which are
then discharged with a floating-point arithmetic solver and~\citeN{Chiang2014}
developed a guided search to find inputs which maximize errors.~\citeN{Lam2013}
uses instrumentation to detect cancellation and thus loss of precision.~\citeN{Ivancic2010}
combines abstract interpretation with model checking to
check the stability of programs, tracking one input at a time.~\citeN{Majumdar2010}
uses concolic execution to find two sets of inputs which
maximize the difference in the outputs. These approach are based on testing,
however, and cannot prove sound bounds.

It is natural to use the Jacobian for sensitivity analysis. Related to our
work is a proof framework using this idea for showing programs robust in the
sense of k-Lipschitz continuity~\cite{Chaudhuri2011}. Note, however, that
our approach does not require programs to be continuous.
~\citeN{Gazeau2012} relaxes the strict definition of robustness to programs with specified
uncertainties and presents a framework for proving while-loops with a
particular structure robust. Our work follows the philosophy of these
approaches in leveraging Jacobians of program paths,
yet we explicitly incorporate the handling of roundoff errors
in a fully automated system.

%% file: conclusion.tex
\section{Conclusion}

We believe that numerical errors, such as roundoff errors, should not be an
afterthought and that  programming language support is needed and possible to
help scientists write numerical code that does what it is expected to do. To
this end, we presented, on one hand, a real-valued specification language with
explicit error annotations from which our tool Rosa synthesizes finite-precision
code that fulfills the given specification. On the other hand, we
presented a set of techniques based on unified principles which provides
automated, efficient, static error analysis which is crucial towards making
such a compiler practical. We have extensively evaluated these techniques
against state-of-the-art tools and we believe they represent an interesting
compromise between accuracy and efficiency.